# HYBRID LS-LMMSE CHANNEL ESTIMATION Technique for LTE Downlink Systems

Abdelhakim Khlifi[1] and Ridha Bouallegue[2]

[1]National Engineering School of Tunis, Tunisia
abdelhakim.khlifi@gmail.com
[2]Sup'Com, Tunisia
ridha.bouallegue@gmail.com

## ABSTRACT

In this paper, we propose to improve the performance of the channel estimation for LTE Downlink systems under the effect of the channel length. As LTE Downlink system is a MIMO-OFDMA based system, a cyclic prefix (CP) is inserted at the beginning of each transmitted OFDM symbol in order to mitigate both inter-carrier interference (ICI) and inter-symbol interference (ISI). The inserted CP is usually equal to or longer than the channel length. However, the cyclic prefix can be shorter because of some unforeseen channel behaviour. Previous works have shown that in the case where the cyclic prefix is equal to or longer than the channel length, LMMSE performs better than LSE but at the cost of computational complexity .In the other case, LMMSE performs also better than LS only for low SNR values. However, LS shows better performance for LTE Downlink systems for high SNR values. Therefore, we propose a hybrid LS-LMMSE channel estimation technique robust to the channel length effect. MATLAB Monte –Carlo simulations are used to evaluate the performance of the proposed estimator in terms of Mean Square Error (MSE) and Bit Error Rate (BER) for 2x2 LTE Downlink systems.

## KEYWORDS

*LTE, MIMO, OFDM, cyclic prefix, channel length, LS, LMMSE*

## 1. INTRODUCTION

To serve the massive growth in demand for mobile broadband services, a high speed data access is required. A lot of researches are done in order to improve high system capacities. Among them, the multiple-input multiple-output (MIMO) represents the most interessant research results. The researches based on MIMO technologies have leading to improve high system capacity without additional bandwidth [1].Multipath propagation causing selective frequency channels may causes serious problems for mobile .Therefore, Multicarrier modulation (MC), especially Orthogonal Frequency Division Multiplexing (OFDM) [2]. This kind of multicarrier modulation has been receiving growing interest in recent years as a solution to combat the effect of frequency selectivity of wireless channels due to simplified equalization in the frequency domain [3]. MIMO is associated with orthogonal frequency-division multiplexing (OFDM) technique. OFDM consists of converting a frequency-selective fading channel into parallel flat-fading sub-channels. The propagation over the radio-frequency channel is characterized by a spread of the signal in time due to multipath propagation and in frequency due to the Doppler effect. This time and frequency dispersion leads to the loss of orthogonality. The transmitted signal will be affected by inter-symbol (ISI) and inter-channel (ICI) interferences. Therefore, a guard time called cyclic prefix CP is added at the beginning of each OFDM symbol in order to mitigate ICI and ISI. Usually, the inserted cyclic prefix is equal to or longer than to the channel [4].

LTE (Long Term Evolution) is the next generation MIMO-OFDM based system. In this paper, we interest to LTE Downlink systems. LTE Downlink system adopts Orthogonal Frequency

Division Multiple Access (OFDMA) as a access technique in the Downlink. LTE Downlink system is MIMO-OFDMA based system which delivers high data rates of up to 100Mbps for 2x2 MIMO systems.

Channel estimation is critical in the design of MIMO-OFDM systems. Many works have studied the performance of channel estimation techniques for MIMO-OFDM systems such as in [3] [4]. However, most of these research works assume that the CP length is equal or longer than the channel length. However, it is also important to study the performance of MIMO-OFDM systems in the case where the cyclic prefix can be shorter than channel length because of some unforeseen channel behaviour. In this case, the task of channel estimation will be more difficult because of the introduction of both ICI and ISI. Equalization techniques were proposed in [8] [9] in order to reduce the effect of ISI and ICI.

The performance evaluation of the two estimators for LTE systems was discussed in many articles e.g [10] [11] The performance of LS and LMMSE channel estimation techniques for LTE Downlink systems under the effect of the channel length was studied in [12].

In this paper, we propose a hybrid LS-LMMSE channel estimation technique which is robust to the channel length effect.

In the rest of the paper, we give a description of LTE Downlink system in section II. A LTE MIMO-OFDM system model is given in section III. The LS and the LMMSE channel estimation techniques are discussed and a description of the proposed hybrid technique is given in section IV with simulation results for its performance under the channel length effect given in section V. Conclusion is given in the last section.

## 2. LTE DOWNLINK PHYSICAL LAYER

Figure 1 illustrates the structure of the LTE radio frame. The duration of one LTE radio frame is 10 ms. It is composed of 20 slots of 0.5 ms. Each two consecutive time slots are combined as one sub-frame and so LTE radio frame contains 10 sub-frames.

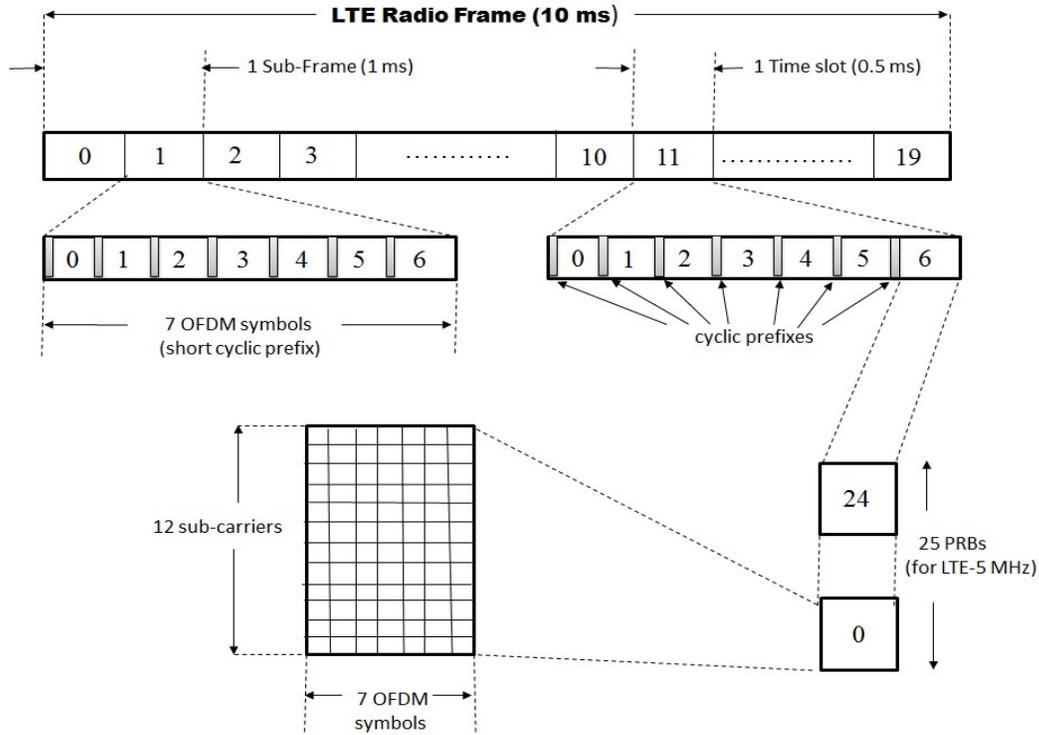

Figure 1: LTE radio Frame structure

The number of OFDM symbols in one slot depends on the chosen length of CP and can be either 6 symbols (for long CP) or 7 symbols (for short CP). The total number of available subcarriers depends on the overall transmission bandwidth of the system as described in table 1. A Physical Resource Block (PRB) is defined as 12 consecutive subcarriers for one slot. The subcarriers are spaced by 15 kHz from each other. Therefore, each PRB occupies a bandwidth of 180 kHz (12 x 15 kHz).A PRB consists of 84 resource elements (12 subcarriers during 7 OFDM symbols) or 72 resource elements (12 subcarriers during 6 OFDM symbols).

The LTE specifications [5] define parameters for system bandwidths from 1.25 MHz to 20 MHz as shown in Table 1

Table 1: LTE Downlink parameters

| Transmission Bandwitdh (MHz) | 1.25 | 2.5 | 5 | 10 | 15 | 20 |
|---|---|---|---|---|---|---|
| Sub-frame duration (ms) | 0.5 | | | | | |
| Sub-carrier spacing (kHz) | 15 | | | | | |
| Physical resource block (PRB) bandwidth (kHz) | 180 | | | | | |
| Number of available PRBs | 6 | 12 | 25 | 50 | 75 | 100 |
| Sampling Frequency (MHz) | 1.92 | 3.84 | 7.68 | 15.36 | 23.04 | 30.72 |
| FFT size | 128 | 256 | 512 | 1024 | 1536 | 2048 |
| Number of occupied sub-carriers | 76 | 151 | 301 | 601 | 901 | 1201 |

## 3. LTE DOWNLINK SYSTEM MODEL

LTE Downlink system is a MIMO-OFDM based system. The system model is given in Figure 2. We consider a MIMO system with $M_T$ transmit antennas and $M_R$ antennas.

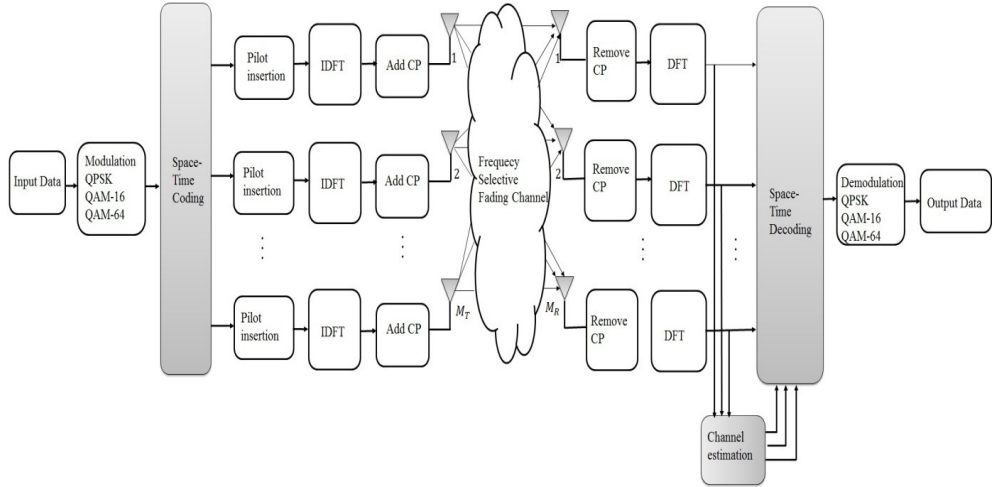

Figure 2: MIMO-OFDM system

LTE Downlink systems employ OFDMA scheme for multiple access. OFDMA is a multiple access scheme based OFDM modulation technique. We consider a OFDM system with $N_{Bw}$ carriers occupying the bandwidth. OFDMA allocates a number of time/frequency resources to different users. Its objective is to split the data stream to be transmitted onto narrowband orthogonal subcarriers by the means of the inverse discrete Fourier Transform (IDFT) operation, which allows for an increased symbol period. A cyclic prefix with the length of $L_{CP}$ is appended at the beginning of each OFDM symbol. The CP consists of a repetition of the last part of an OFDM symbol. In order to avoid degradations due to inter-symbol interference (ISI) and inter-carrier interference (ICI), the inserted CP is generally equal or longer than the maximum excess delay of the channel.

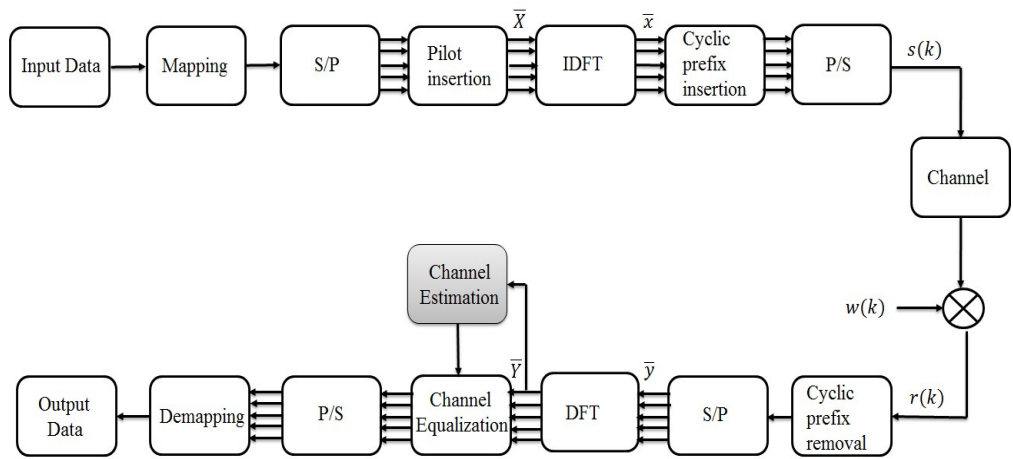

Figure 3: Baseband OFDM system

Each OFDM symbol is transmitted over frequency-selective fading MIMO channels assumed independents of each other. The multipath fading channel can be modeled as a FIR (Finite Impulse Response) with *L* taps for each channel.

$$g(t, \tau) = \sum_{l=0}^{L-1} g_l(t)\delta(t - \tau_l) \quad (1)$$

$g_l(t)$ and $\tau_i$ are respectively the impulse response and the multipath delays of the channel. It is sufficient to consider in our system model only a single transmit and a single receive antenna.

We note
$\overline{X} = [X_0 X_1 ... X_{N_{BW}-1}]^T$ : The input data of IDFT block at the transmitter.
$\overline{Y} = [Y_0 Y_1 ... Y_{N_{BW}-1}]^T$ : The output data of DFT block at the receiver.
$\overline{g} = [h_0 h_1 ... h_{N_{BW}-1}]^T$ : The sampled channel impulse response
$\overline{w} = [w_0 w_1 ... w_{N_{BW}-1}]^T$ : The sampled additive noise

Let us define $\underline{X} = diag(\overline{X})$ and the input *N*-DFT matrix as:

$$\underline{F} = \begin{pmatrix} f_N^{0,0} & \cdots & f_N^{0,N_{BW}-1} \\ \vdots & \ddots & \vdots \\ f_N^{N-1,0} & \cdots & f_N^{N-1,N_{BW}-1} \end{pmatrix} \quad (1)$$

where

$$f_N^{l,k} = \frac{1}{\sqrt{N}} e^{-j2\pi lk/N} \quad (2)$$

We define also the frequency channel given by:

$$\overline{H} = DFT_N(\underline{g}) = \underline{F}\overline{g} \quad (3)$$

and the noise in the frequency domain given by:

$$\overline{W} = DFT_N(\overline{w}) = \underline{F}\overline{w} \quad (4)$$

The received OFDM symbol, after removing the CP and performing the DFT, at one receive antenna can be written as:

$$\begin{aligned}\overline{Y} &= DFT_N(IDFT_N(\overline{X}) \otimes \overline{g} + \overline{W}) \\ &= \underline{X}\underline{F}\overline{g} + \overline{W} = \underline{X}\overline{H} + \overline{W}\end{aligned} \quad (5)$$

$\overline{Y}$ represents the received signal vector; $\underline{X}$ *is* a diagonal matrix containing the reference symbols; $\overline{H}$ is a channel frequency response vector and $\overline{W}$ is an additive complex-valued white Gaussian noise vector with zero mean and variance $\sigma_w^2$. The noise is assumed to be independent of the transmitted signals.

## 4. CHANNEL ESTIMATION

Channel estimation has important impact on the receiver performance. In LTE systems, channel estimation is performed based on the reference signals [5]. The reference symbols are placed in the first OFDM symbol and the fifth OFDM symbol of every time slot when short CP is used while they are placed in the first and the fourth OFDM of every time slot when long CP is used. Figure 3 shows reference signal pattern for two antennas.

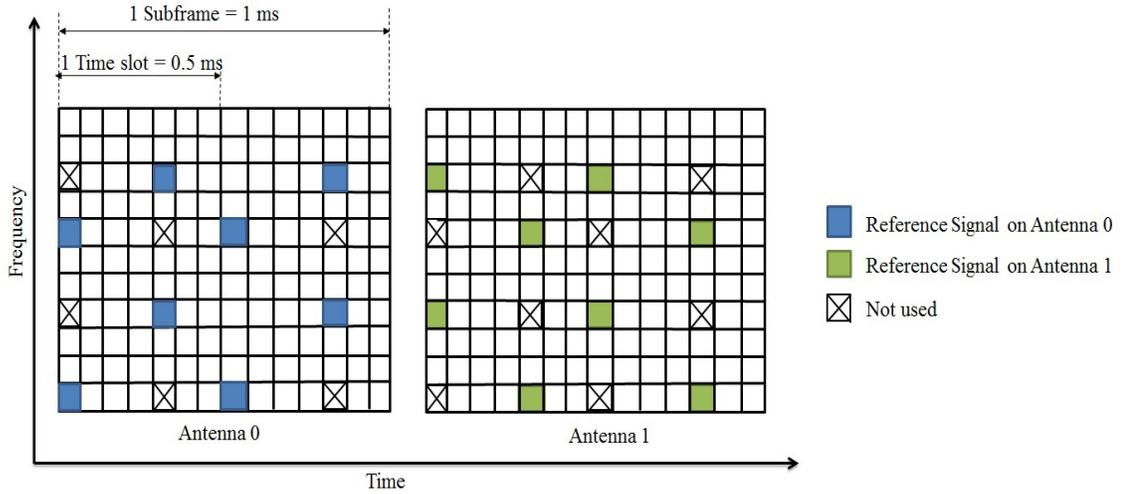

Figure 4: Reference signal pattern for two antennas [5]

From (1), the received pilot signals can be written as:

$$\overline{Y}_p = \underline{X}_p \overline{H}_p + \overline{W}_p \qquad (6)$$

$(.)_p$ denotes positions where reference signals are transmitted.
The task now is to estimate the channel responses given the pilot signals $X_P$ and the received and the received signals $Y_P$. In this paper, two linear channel estimators are studied: the Least Square (LS) and the Linear Minimum Mean Square Error (LMMSE).

### 4.1 Least Square LS

The LS estimate of such system is obtained by minimizing the square distance between the received signal $\overline{Y}$ and the original signal $\underline{X}$.

The least square estimates (LS) of the channel at the pilot subcarriers given in (6) can be obtained as [6]:

$$\hat{H}_P^{LS} = (\underline{X}_P)^{-1} \overline{Y}_P \qquad (7)$$

.
The LS estimators are known by its very low complexity because they not need the statistic information of channel.

**4.2 Linear Mean Minimum Square Error LMMSE**

The LMMSE channel estimation employs the channel statistics to minimize the MSE estimate of the channel responses given in (6) is [7]:

$$H_p^{LMMSE} = R_{HH_p}\left(R_{H_pH_p} + \sigma_u^2(XX^H)^{-1}\right)^{-1} \hat{H}_p^{LS} \quad (7)$$

$R_{HH_p}$ is the cross-correlation matrix between all subcarriers and the subcarriers with reference signals. $R_{H_pH_p}$ is the autocorrelation matrix of the subcarriers with reference signals. The LMMSE estimators are known by their high complexity due to the inversion matrix lemma given in (7). The problem with the LMMSE estimator that it needs to apply the inversion $(XX^H)^{-1}$ any time the data in changes. In order to reduce the complexity of the LMMSE estimator, the term $(XX^H)^{-1}$ will be replaced by its expectation $E\{(XX^H)^{-1}\}$ in (7). If we assume the same signal constellation on all tones and equal probability on all constellation points, we obtain $E\{(XX^H)^{-1}\} = E\{|1/X_k|^2\}$. By defining the average SNR as $E[|X_k|^2]/\sigma_w^2$, the simplified LMMSE estimator becomes [7]:

$$\hat{H}_p^{LMMSE} = R_{HH_p}\left(R_{H_pH_p} + \frac{\beta}{SNR}I_P\right)^{-1} \hat{H}_p^{LS} \quad (8)$$

where $\beta$ is scaling factor depending on the signal constellation (i.e $\beta = 1$ for QPSK and $\beta = 17/9$ for 16-QAM) and $I_P$ is the identity matrix.

**4.3 Hybrid LS-LMMSE Channel estimation technique**

The performance of LS and LMMSE channel estimation techniques for LTE Downlink systems under the channel effect was studied in [12]. Simulation results have shown that, in the case where the cyclic prefix is longer than the channel, the LMMSE estimation technique is better than the LS estimator. The LMMSE estimator offers better performance than the LS estimator for LTE Downlink systems but at the cost of high complexity due to the inversion lemma. However, in other case; even the cyclic prefix is shorter than the channel length; LMMSE shows also better performances than LS for LTE Downlink systems but only for low SNR values. For high SNR values, LMMSE loses its performance in terms of MSE and in other hand; LS estimator seems to perform better than LMMSE for this range of SNR values.

Therefore, we propose a hybrid LS-LMMSE estimation technique robust to the channel effect. The proposed technique will act depending on the channel length. In the case where the cyclic prefix is equal to or longer than the channel length, the hybrid LS-LMMSE algorithm will apply directly the LMMSE channel estimation technique. In other hand, the case where the inserted cyclic prefix is shorter than the channel length, the proposed hybrid will act now depending on the received SNR value: when the SNR value is low, the hybrid LS-LMMSE algorithm will choose to apply the LMMSE estimator. However, when the received SNR value is considered high, the proposed algorithm will switch to the LS estimator. It is clear that in this case where the channel length exceeds the cyclic prefix length, the hybrid LS-LMMSE channel estimator shows its true efficiency.

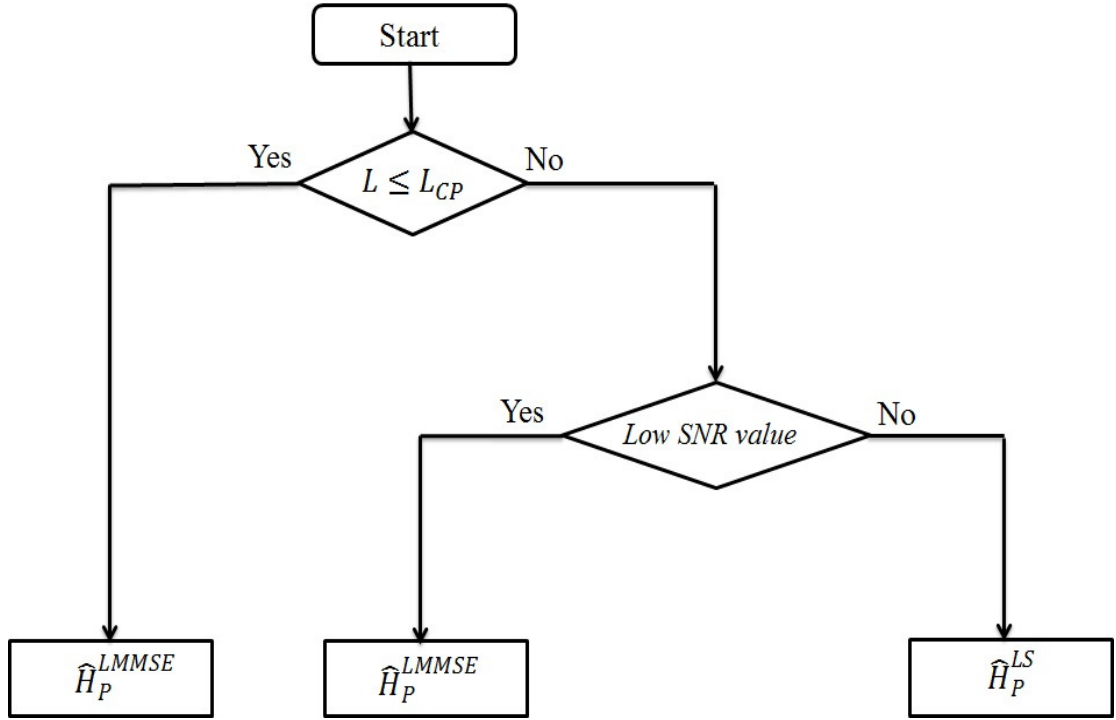

Figure 5: Hybrid LS-LMMSE channel estimation algorithm

## 5. SIMULATION RESULTS

Simulation studies are conducted to investigate the performance of the proposed hybrid LS-LMMSE estimation technique for LTE Downlink systems under the effect of the channel length. We consider a LTE-5 MHz Downlink system with 2 transmit and 2 receive antennas. The number of used subcarriers is set to 300. The length of cyclic prefix is $L_{CP} = 16$. The transmitted signals are quadrature phase-shift keying (QPSK) modulated. 100 radio frames are sent through a frequency-selective channel. The frequency-selective fading channel responses are randomly generated with a Rayleigh probability distribution. Table 2 gives a summary of the simulation parameters.

Table 2: Simulation Parameters

| Simulation Parameters | |
|---|---|
| LTE Bandwidth | 5 MHz |
| Number of used subcarriers | 300 |
| Cyclic prefix length | 16 |
| Number of transmitted Frames | 100 |
| Number of transmit antenna | 2 |
| Number of Receive antenna | 2 |
| Modulation scheme | QPSK |
| Channel Model | Rayleigh |

## 5.1 Case with $L \leq L_{CP}$

In this case, the cyclic prefix is longer than the channel which means that ISI and ICI are completely suppressed. The hybrid LS-LMMSE estimation algorithm will apply directly the LMMSE estimation technique. The proposed technique gives the same performance results as in [12]. The LMMSE estimator performs better than the LS estimator but at the cost of computational complexity. Figure 6 and Figure 7 show respectively the performance of the hybrid LS-LMMSE estimator in terms of BER and MSE for $L$=6 and $L$=10.

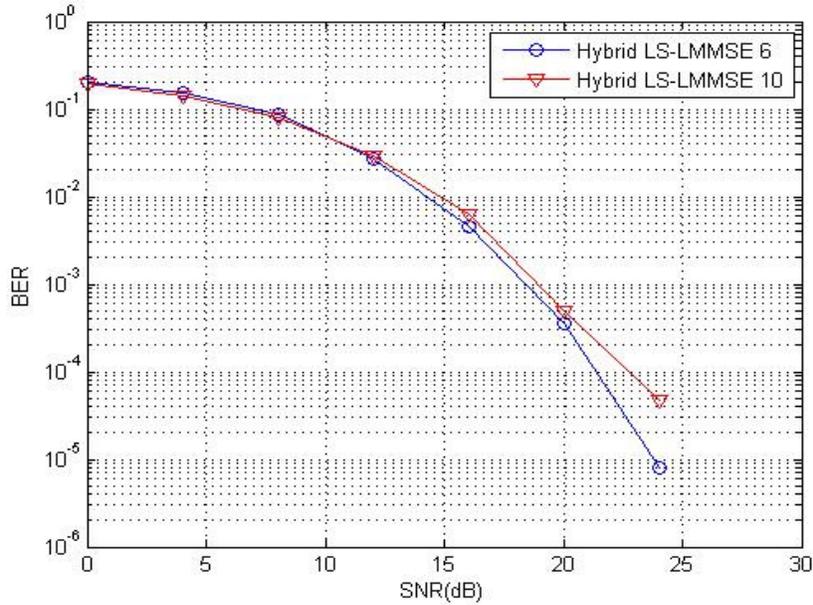

Figure 6: BER versus SNR for $L = 6$ and $L = 10$

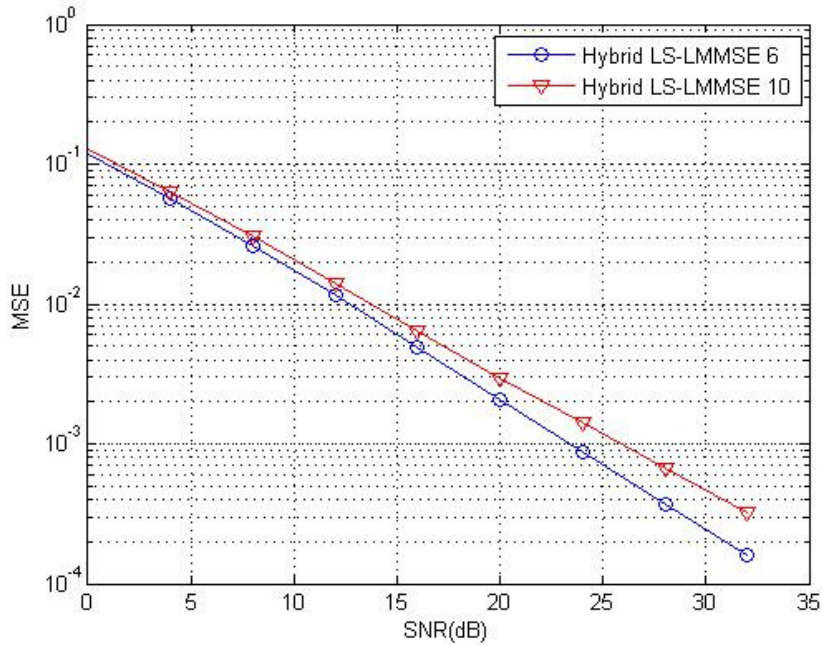

Figure 7: BER versus SNR for $L = 6$ and $L = 10$

## 5.2 Case with $L \gg L_{CP}$

In this case, the hybrid LS-LMMSE shows its true efficiency for the performance of LTE Downlink systems. The inserted cyclic prefix is shorter than the channel length and this can be due of some unforseen behaviour oh the channel. That will cause the introduction of ISI and ICI. Figure 8 and Figure 9 show the performance of the proposed technique in terms of BER. Figure 10 and Figure 11   show the performance of the proposed technique in terms of MSE.

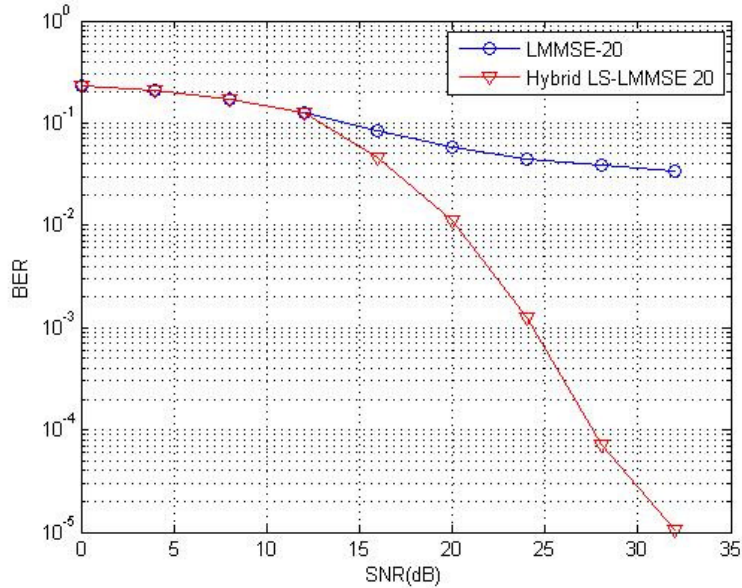

Figure 8: BER versus SNR for $L = 20$

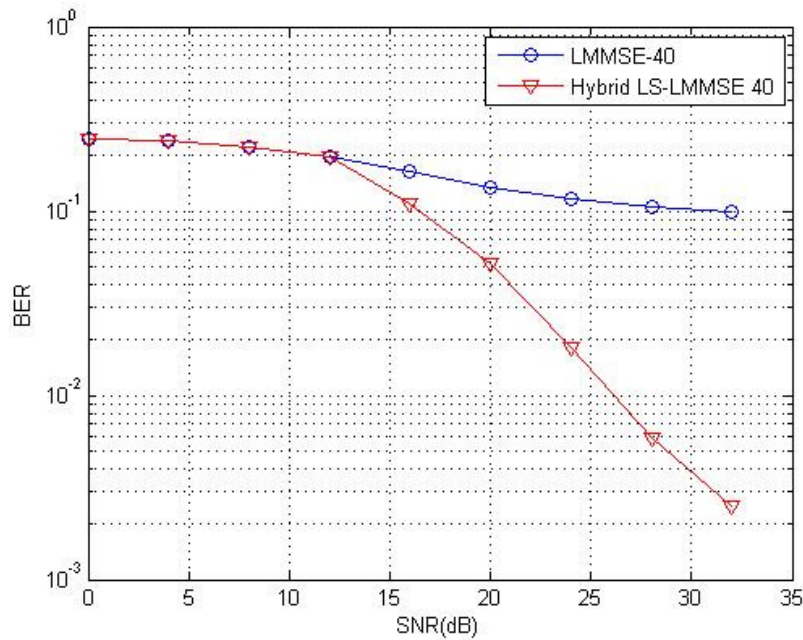

Figure 9: BER versus SNR for $L = 40$

Simulation results show that the hybrid LS-LMMSE estimator performs better the LMMSE estimator especially for high SNR values. For this range of SNR values, we see a large difference between the two estimators and this proves well the efficiency of the proposed technique.

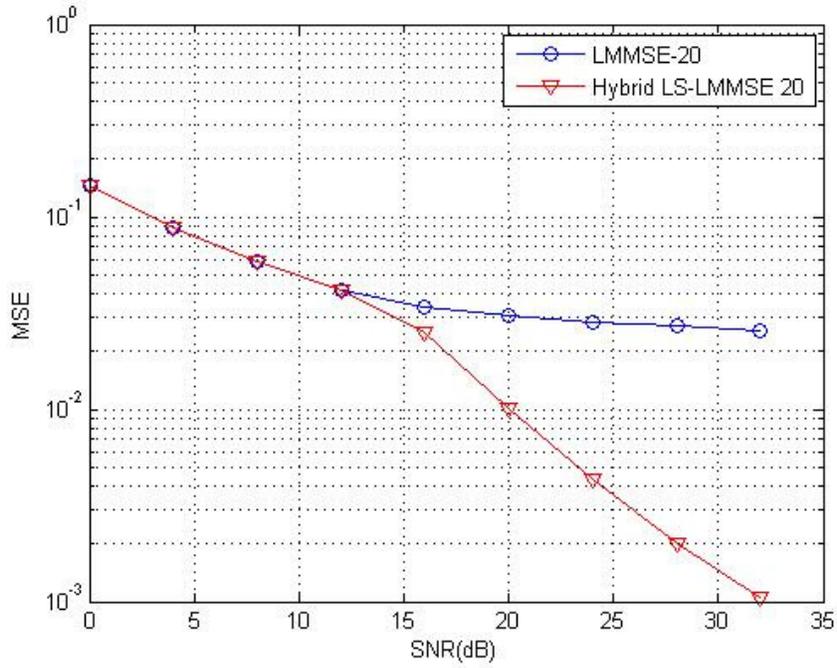

Figure 10: MSE versus SNR for $L = 20$

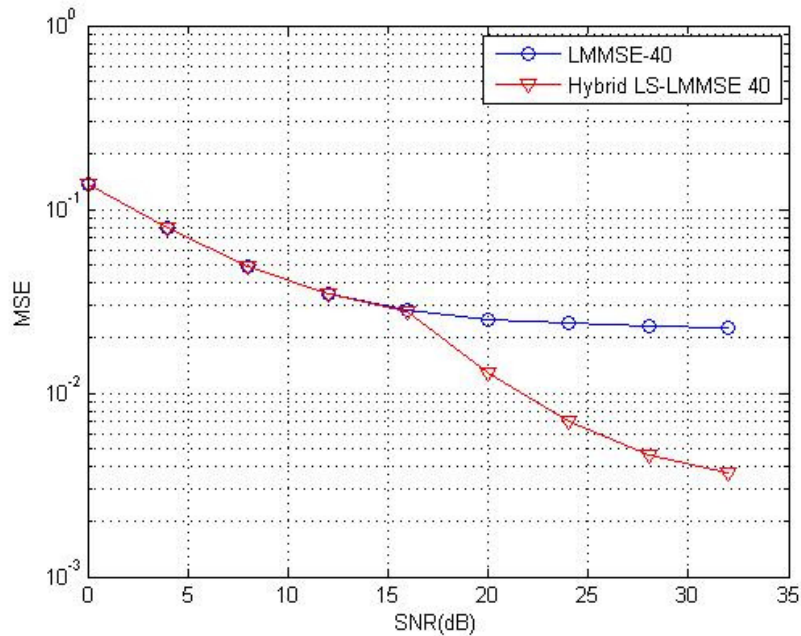

Figure 11: MSE versus SNR for $L = 40$

## 6. Conclusion

Channel estimation is a challenging problem in wireless systems. Most of the research works for MIMO-OFDM systems are based on the assumption that the inserted cyclic prefix at the beginning of each transmitted OFDM symbol is equal to or longer than the channel length in order to suppress ICI and ISI. However, it is important also to study the other case where the CP length can be shorter than the channel length because of some unforeseen behaviour of the channel. Previous works have shown that the LMMSE performs better than LS estimator in the case where the CP length is equal to or longer than the channel length but at the cost of the complexity due to its dependence to the channel and noise statistics. In the other case, LMMSE shows better performance only for low SNR values and begins to lose its performance for higher SNR values. In other hand, LS looks to perform better than LMMSE in this range of SNR values. In this paper, we propose a hybrid LS-LMMSE estimation technique able to perform LTE Downlink systems under the channel length effect. Simulations results have shown the efficiency of the proposed technique and especially for the case where the channel length exceeds the cyclic prefix. In this case, the hybrid technique applies the LMMSE estimation technique only for low SNR values. For high SNR values, the proposed technique switches to the LS estimator.

**Authors**

**Abdelhakim KHLIFI** was born in Düsseldorf, Germany, on January 07, 1984. He graduated in Telecommunications Engineering, from National Engineering School of Gabès in Tunisia, July 2007.In June 2009, he received the master's degree of research in communication systems of the School of Engineering of Tunis ENIT. Currently he is a Ph.D student at the National School of Engineering of Tunis. His research spans radio channel estimation in LTE MIMO-OFDM systems

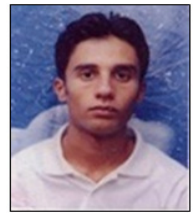

**Ridha BOUALLEGUE** is Professor at the National Engineering School of Tunis, Tunisia (ENIT). He practices at the Superior School of communication of Tunis (Sup'Com). He is founding in 2005 and Director of the Research Unit "Telecommunications Systems: 6'Tel@Sup'ComˇT. He is founding in 2005, and Director of the National Engineering School of Sousse. He received his PhD in 1998 then HDR in 2003. His research and fundamental development, focus on the physical layer of telecommunication systems in particular on digital communications systems, MIMO, OFDM, CDMA, UWB, WiMAX, LTE, has published 2 book chapters, 75 articles in refereed conference lectures and 15 journal articles (2009).

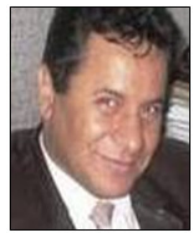